\documentclass[showpacs,prd,aps,superscriptaddress]{revtex4}
\usepackage{amsmath,amssymb,bm}
\usepackage{epsfig}
\usepackage{graphicx}
\usepackage{amsfonts}
\usepackage{epstopdf}
\usepackage{color}

\renewcommand\a{\alpha}
\renewcommand\b{\beta}
\renewcommand\d{\delta}

\renewcommand\l{\lambda}
\renewcommand\r{\rho}

\renewcommand\c{\chi}
\renewcommand\j{\psi}
\renewcommand\o{\omega}

\newcommand\g{\gamma}

\newcommand\m{\mu}
\newcommand\n{\nu}

\newcommand\p{\pi}
\newcommand\h{\theta}
\newcommand\s{\sigma}
\newcommand\f{\phi}

\newcommand\ve{\varepsilon}

\newcommand\G{\Gamma}

\newcommand{\fig}[1]{Fig.~\ref{#1}}
\newcommand{\eq}[1]{Eq.~(\ref{#1})}

\newcommand\lb{\left(}
\newcommand\rb{\right)}
\newcommand\ls{\left[}
\newcommand\rs{\right]}
\newcommand\lc{\left\{}
\newcommand\rc{\right\}}
\newcommand{\lan}{\langle}
\newcommand{\ran}{\rangle}

\newcommand\ra{\rightarrow}

\newcommand{\non}{\nonumber\\}
\newcommand\pt{\partial}

\newcommand{\Tr}{{\rm Tr}}

\newcommand{\bp}{{\vec p}}
\newcommand{\bk}{{\vec k}}
\newcommand{\bq}{{\vec q}}

\newcommand{\jb}{{\bar \j}}

\renewcommand{\part}{{\rm part}}

\renewcommand{\vec}{\boldsymbol}

\begin{document}

\title{Chiral electric separation effect in the quark-gluon plasma}

\author{Yin Jiang}
\email{jiangyin@indiana.edu}
\affiliation{ Physics Department and Center for Exploration of Energy and Matter,
Indiana University, 2401 N Milo B. Sampson Lane, Bloomington, IN 47408, USA.}

\author{Xu-Guang Huang}
\email{huangxuguang@fudan.edu.cn}
\affiliation{Physics Department and Center for Particle Physics and Field Theory, Fudan University, Shanghai 200433, China.}

\author{Jinfeng Liao}
\email{liaoji@indiana.edu}
\affiliation{ Physics Department and Center for Exploration of Energy and Matter,
Indiana University, 2401 N Milo B. Sampson Lane, Bloomington, IN 47408, USA.}
\affiliation{RIKEN BNL Research Center, Bldg. 510A, Brookhaven National Laboratory, Upton, NY 11973, USA.}

\date{\today}

\begin{abstract}
In this paper we introduce and compute a new transport coefficient for the quark-gluon plasma (QGP) at very high temperature. This new coefficient $\sigma_{\chi e}$, the CESE (Chiral Electric Separation Effect) conductivity, quantifies the amount of axial current $\vec J_A$ that is generated in response to an externally applied electric field $e\vec E$: $\vec J_A = \s_{\chi e} (e\vec E)$. Starting with rather general argument in the kinetic theory framework, we show how a characteristic structure $\s_{\chi e}\propto \mu \mu_5$ emerges, which also indicates the CESE as an anomalous transport effect occurring only in a parity-odd environment with nonzero axial charge density $\mu_5\neq 0$. Using the Hard-Thermal-Loop framework the CESE conductivity for the QGP is found to be $\s_{\chi e} = (\#) T\frac{\Tr_{\rm f}Q_eQ_A}{g^4\ln(1/g)}  \frac{\m\m_5}{T^2}$ to the leading-log accuracy with the numerical constant $(\#)$ depending on favor content, e.g. $(\#)=14.5163$ for $u,d$ light flavors.
\end{abstract}
\maketitle

\section{Introduction}

In high energy heavy ion collisions, it is believed that a new form of strong interaction matter, the quark-gluon plasma (QGP), is created and observed. While normal nuclear matter has its smallest building blocks ---  the quarks and gluons --- deeply confined inside hadrons, when heated to high enough temperature these quarks and gluons get deconfined to form the quark-gluon plasma. The QGP is primordially hot and dense: it once occupied the early Universe at a few microseconds after the Big Bang, and is now the hottest matter created at the Relativistic Heavy Ion Collider (RHIC) and the Large Hadron Collider (LHC).

Transport properties provide unique means of probing the structure and dynamics of matter. Electromagnetic (EM) fields, for example, were used hundreds of  years ago to induce charge transport processes inside normal materials, the responses of which were measured and the interpretations of which significantly advanced our understanding of those condensed matter systems. Important findings such as Ohm's law, the generation of electric current in response to external electric field $\vec J_V = \sigma (e\vec E)$ where $e>0$ is the absolute value of electron charge, were made. More recent examples of probing matter with EM field induced transport include, e.g. the famous quantum hall systems.

For the QGP as a new form of strong interaction matter, its charge transport properties under external (Maxwell) electric and magnetic fields are of great theoretical interest as well as of phenomenological importance. In particular, significant efforts have been made to understand the {\it anomalous transport effects}  in the QGP, with the so-called Chiral Magnetic Effect (CME)~\cite{Kharzeev:2004ey,Kharzeev:2007tn,Kharzeev:2007jp} as a distinguished example. The term ``anomalous'' implies that these effects do not exist in normal circumstance due to symmetry constraint (e.g. parity) but can occur in a parity-odd {\it environment}. In the hot QGP created in heavy ion collisions as well as that occupying early Universe, such an environment (a P-odd bubble of QGP) may indeed exist due to the topological fluctuations in the gluonic configurations of Quantum Chromodynamics (QCD) (see e.g. discussions in \cite{Kharzeev:2007tn,Zhitnitsky:2012im}). With chiral symmetry restoration at high temperature the light flavor quarks are approximately chiral in QGP and through the famous triangle anomaly the parity-oddness from QCD topological fluctuations gets translated into an imbalance of quark chirality (i.e. difference between number of right-handed and left-handed quarks), usually parameterized by a nonzero axial chemical potential $\mu_5$. The CME predicts the generation of an electric current $\vec J_V$ in response to an external magnetic field $\vec B$, $\vec J_V \propto \mu_5 (e\vec B)$. The coefficient of proportionality (CME conductivity) was computed and found to be fixed by axial anomaly uniquely and universal from weak to strong coupling if the dynamical gauge bosons are absent~\cite{Fukushima:2008xe,Yee:2009vw,Rebhan:2009vc,Gynther:2010ed,Sadofyev:2010pr,Kalaydzhyan:2011vx}. When the dynamical gauge bosons are included in the calculation, however, the CME conductivity may receive radiative corrections (see e.g.~\cite{Jensen:2013vta,Gorbar:2013upa}). The case of an external electric field $\vec E$ was  investigated only very recently and a new effect, the Chiral Electric Separation Effect (CESE), was found in \cite{Huang:2013iia}. The CESE predicts the generation of an axial current $\vec J_A$ in response to the electric field, $\vec J_A \propto \mu \mu_5 (e\vec E)$ (where $\mu$ is the chemical potential corresponding to certain vector charge density). Interesting collective excitations arising from interplay between vector and axial currents in external $\vec E$ and/or $\vec B$ fields have also been studied and nontrivial modes such as Chiral Magnetic Wave (CMW)~\cite{Kharzeev:2010gd,Burnier:2011bf} and Chiral Electric Wave (CEW)~\cite{Huang:2013iia,Pu:2014fva} have been identified.
The CMW is based on the mutual induction of $\vec J_{V,A}$ via the CME and another effect known as Chiral Separation Effect (CSE, which induces an axial current along magnetic field)~\cite{son:2004tq,Metlitski:2005pr} with both of them propagating in an entangled manner. Finally as pointed out first in \cite{Kharzeev:2007tn}, in a relativistically rotating fluid a quantum particle feels a similarity between fluid vorticity and an external magnetic field, and thus a new effect in analogy to CME has been found to predict the generation of a vector current along the vorticity $\vec J_V \propto \mu\mu_5 \omega$ known as Chiral Vortical Effect (CVE)~\cite{Kharzeev:2007tn,Son:2009tf,Kharzeev:2010gr}. For recent reviews on such anomalous transport effects, see e.g.~\cite{Kharzeev:2013ffa,Liao:2014ava,Bzdak:2012ia}.

An experimental observation of any such anomalous transport effect in e.g. heavy ion collisions would provide indications of the manifestation of QCD topology and anomaly as well as possible ``environmental symmetry violation'' in primordial matter. It is thus of fundamental significance. In recent years a lot of experimental efforts have been invested to look for one or more of the anomalous transport effects in heavy ion collisions at RHIC and the LHC~\cite{STAR:2009uh,Ajitanand:2010rc,Abelev:2012pa,Adamczyk:2014mzf,Wang:2012qs}. Charge-dependent azimuthal correlations have been measured across a wide span of collision beam energy to search for a charge separation across the reaction plane (i.e. an out-of-plane charge dipole) possibly induced by CME~\cite{Kharzeev:2004ey,Kharzeev:2007jp}. The charge-dependent splitting of pion elliptic flow has also been measured with very encouraging features in line with a charge quadruple distribution of QGP induced by CMW~\cite{Burnier:2011bf,Wang:2012qs}. A new measurement of a possible baryon separation across the reaction plane has also been released by the STAR collaboration~\cite{Zhao:2014}, which might be induced by CVE. Many works have been done in order to fully and accurately extract the interpretations of these data and their implications for the search of anomalous transport effects~\cite{Bzdak:2009fc,Hongo:2013cqa}. At this point the status of such search is not conclusive yet (for reviews see e.g.~\cite{Kharzeev:2013ffa,Liao:2014ava,Bzdak:2012ia}), with more detailed measurements, more sophisticated modelings, as well as more precise theoretical determinations of pertinent transport coefficients badly needed.

 This paper will focus on studying the newly introduced Chiral Electric Separation Effect in the QGP. Specifically, we aim to quantify a new transport property, the CESE conductivity, of the QGP in a high enough temperature regime where a perturbative framework can be justified. Previous works have computed this conductivity for a QED plasma~\cite{Huang:2013iia} and for strongly coupled holographic gauge plasma~\cite{Pu:2014cwa}. The precise value of CESE conductivity for QGP, however, is not calculated yet to the best of our knowledge. It is the purpose of this paper to determine this quantity. We will first use the kinetic theory framework to develop an intuitive picture of the emergence of the CESE and point out the general structure of the CESE conductivity. Next we use the Hard-Thermal-Loop (HTL) framework to compute the CESE conductivity to the leading-log accuracy for high temperature QGP.  Summary of our results and physical discussions will given in the end.

\section{CESE in kinetic theory}

Kinetic theory provides a natural and simple framework to intuitively understand the current generation as linear response to externally applied fields. Very recently a number of studies~\cite{Son:2012wh,Stephanov:2012ki,Gao:2012ix,Satow:2014lva} have used kinetic framework to  understand the various anomalous transport effects mentioned above. Here let us use this approach to show in general how the CESE arises in a plasma containing charged chiral fermions. In kinetic theory, we use a phase space distribution function $f_{a,s}(\vec x,\vec p; t)$ to describe each species of the fermions, with indices defined as follows: $f_{a,s}$ with $a=\pm1$ for positively/negatively charged and $s=\pm1$ for right-handed/left-handed fermions, respectively. The distribution function evolves according to the Boltzmann equation
\begin{eqnarray}
\label{boltzmann}
\partial_t f_{a,s} + \vec v \cdot \nabla_{\vec x} f_{a,s} + \vec F_{a,s} \cdot \nabla_{\vec p} f_{a,s} = C[f_{a,s},...],
\end{eqnarray}
where the $\vec F_{a,s}$ on the left-hand side (LHS) is the external force acting on the fermions and the collision kernel on the right-hand side (RHS) is a functional depending on the distributions of all particle species involved in the microscopic interactions with the fermions (including those beyond the fermions we consider, e.g. gluons).

At this point, some comments are in order. For a plasma containing chiral fermions, the kinetic equation (\ref{boltzmann}) is not complete because it lacks the contributions from the Berry curvatures~\cite{Son:2012wh,Stephanov:2012ki,Gao:2012ix}. The presence of the Berry curvatures will modify the identifications of $\vec v$ with $\dot{\vec x}$ and $\vec F_{a,s}$ with $\dot{\vec p}$ and thus introduce new terms into the kinetic equation. These new terms encode the axial anomaly into the kinetic equation and are responsible to the kinetic description of the CME and CSE. However, for our purpose, these Berry curvature terms do not play a role because we will consider a plasma in a homogenous electric field with no external magnetic field being present. Thus, according to Refs.~\cite{Son:2012wh,Stephanov:2012ki,Gao:2012ix}, all the Berry curvature induced terms disappear. This illustrates that the CESE, although arising only in P-odd environments, does not originate directly from the axial anomaly.

For demonstration of current generation in the linear response regime, we consider the relaxation time approxiation, i.e. $C \to - (f_{a,s} - f^0_{a,s})/\tau_{a,s}$ with $\tau_{a,s}>0$ the relaxation time for each specific species. Here $f^0_{a,s}$ is the equilibrium distribution while the $\delta f= f-f_0$ is the small deviation from equilibrium due to the externally applied fields. Specifically we consider a homogeneous electric field $\vec E$ thus the force is $\vec F_{\pm,s}= \pm e \vec E$  with $e$ the electric charge (For simplicity we assume all the species to have the same charge number). To the linear order of deviation from equilibrium, it is not difficult to see that a static and homogeneous solution in this case is given by:
\begin{eqnarray}
\delta f_{\pm,s} =  (- \tau_{\pm,s})\, (\pm e \vec E) \cdot \nabla_{\vec p} f^0_{\pm,s}.
\end{eqnarray}
Now one can compute the current generated. Let's first look at the usual electric current, given by:
\begin{eqnarray}
\vec J_V &=& \int_{\vec p} \vec{v} \left[ f_{+,+} + f_{+,-} - f_{-,+} - f_{-,-} \right] = e\vec E  \sum_{a,s} \tau_{a,s}U^0_{a,s},
\end{eqnarray}
where $U^0_{a,s} = \int_{\vec p} \frac{1-(\vec v \cdot \hat E)^2}{p} f^0_{a,s}>0$ and $\int_{\vec p}=\int d^3\vec p/(2\pi)^3$. Note that in the above the equilibrium part does not contribute to the net current. The above relation is of course nothing but the familiar Ohm's law $\vec J_V = \sigma (e\vec E)$ with $\sigma=\sum_{a,s} \tau_{a,s}U^0_{a,s} >0$ the usual electric conductivity.

Now let us examine the axial current, given by:
\begin{eqnarray}
\vec J_A &=& \int_{\vec p} \vec{v} \left[ f_{+,+} + f_{-,+} - f_{+,-} - f_{-,-} \right] \nonumber \\
&=& e\vec E   [ \tau_{+,+}U^0_{+,+} +  \tau_{-,-}U^0_{-,-}   - \tau_{+,-}U^0_{+,-} -\tau_{-,+}U^0_{-,+}   ].
\end{eqnarray}
The above allows an identification of the CESE conductivity $\vec J_A = \sigma_{\chi e} (e\vec E)$ in analog to the electric conductivity:
\begin{eqnarray}
\sigma_{\chi e} =  \tau_{+,+}U^0_{+,+} +  \tau_{-,-}U^0_{-,-}   - \tau_{+,-}U^0_{+,-} -\tau_{-,+}U^0_{-,+}.
\end{eqnarray}
In the above the $\tau_{a,s}$ and $U^0_{a,s}$ are transport and thermal properties of the system evaluated at equilibrium, and are thus dependent on the relevant thermodynamic parameters: temperature $T$, vector chemical potential $\mu$, and axial chemical potential $\mu_5$. On quite general ground one expects the following structure via Taylor expansion at small chemical potentials to quadratic order:
\begin{eqnarray}
 \tau_{a,s} U^0_{a,s} &\equiv& V_{a,s} \nonumber \\
& =& V(T)  {\bigg [} 1 + d_{10}  \frac{a\, \mu}{T}   + d_{01} \frac{s\, \mu_5}{T}   + d_{20} \left(\frac{a\,  \mu}{T}\right)^2 + d_{02} \left(\frac{s\,  \mu_5}{T}\right)^2  + d_{11} \left(\frac{a\, s\, \mu\mu_5}{T^2}\right) {\bigg ]}, \quad
\end{eqnarray}
where $V(T)$ is a function of the temperature only and $d$'s are constant.
With the above expansion it follows immediately that
\begin{eqnarray} \label{eq_conductivities_kinetic1}
\sigma &=& 4V(T) \left[ 1 + d_{20} \left(\frac{ \mu}{T}\right)^2 + d_{02} \left(\frac{ \mu_5}{T}\right)^2 \right],\\
\sigma_{\chi e} &=& 4V(T) \left[  d_{11}\, \left(\frac{\mu}{T}\right)\, \left(\frac{\mu_5}{T}\right) \right].
\label{eq_conductivities_kinetic2}
\end{eqnarray}
A number of comments are in order here:\\
 1) The CESE conductivity $\sigma_{\chi e}$ is nonzero only when both $\mu$ and $\mu_5$ are nonzero, as previously found \cite{Huang:2013iia}.\\
 2) While the normal conductivity $\sigma\geq0$ owing to its dissipative nature (as is evident from the work supplied by external field $\sim \vec J_V \cdot \vec E$), the CESE conductivity $\sigma_{\chi e}$ can be either positive or negative which, at first sight, may suggest a non-dissipative nature of this current. However, on the other hand, $\sigma_{\chi e}$ is time-reversal odd because $\vec J_A$ is time-reversal odd but $\vec E$ is time-reversal even. This time-reversal oddness of $\sigma_{\chi e}$ suggests CESE to be dissipative~\cite{Kharzeev:2011ds}. Thus the dissipation property of CESE is quite nontrivial and deserves a careful analysis. We put such an analysis in the Appendix~\ref{diss}.\\
 3) On the other hand it should also be pointed out that the above derivation suggests both currents to arise from conducting transport in external electric field. \\
 4) It will be shown later via explicit computations that the above structures are true for these conductivities.

\section{CESE conductivity of high temperature quark-gluon plasma}

In this section, we compute the CESE conductivity for QGP at very high temperature. In such temperature regime the QGP is weakly coupled and the Hard-Thermal-Loop (HTL) approach provides a standard framework for calculating various transport coefficients, see e.g.~\cite{Jeon:1994if,Jeon:1995zm,MartinezResco:2000pz,ValleBasagoiti:2002ir,Aarts:2002tn,Aarts:2005vc,Defu:2005hb,Gagnon:2006hi,Gagnon:2007qt,Carrington:2007fp,Arnold:2000dr,Arnold:2003zc,Dobado:2001jf,Aarts:2004sd,Chen:2006iga,Chen:2011km}.  We will extend such HTL-based method to determine the CESE conductivity in a very hot quark-gluon plasma with nonzero vector and axial chemical potentials.

\begin{figure}[!htb]
\begin{center}
\includegraphics[width=5.5cm]{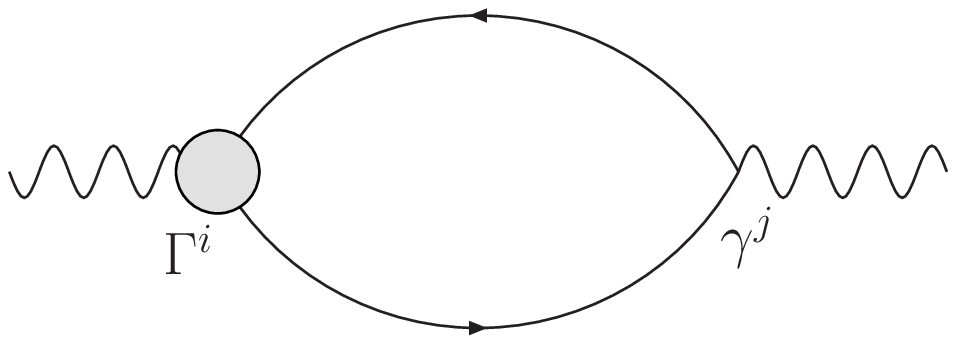} \hspace{0.5cm}
\includegraphics[width=5.5cm]{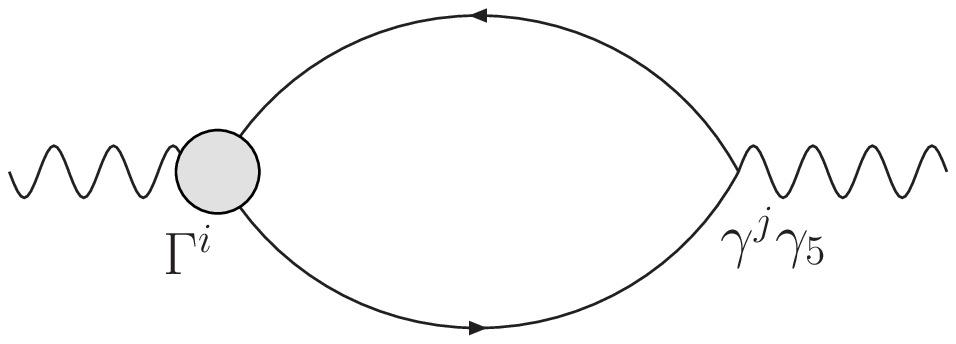}
\caption{The Feynmann diagrams for retarded correlation functions $G_{Ve}^R$ (Left) and $G_{Ae}^R$ (Right).} \label{diag:kubo}
\end{center}
\end{figure}

We consider the QGP in a general situation, with given temperature $T$, $U_V(1)$ vector chemical potential $\mu$, and $U_A(1)$ axial chemical potential $\mu_5$. Furthermore we consider $N_f$ quark flavors, with  $Q_V$ and $Q_A$ the vector and axial charge matrices in flavor space. For example, $Q_B=(1/3)\mathbb{I}$ for baryon current (in $N_c=3$ case), while for charge current $Q_e=Diag(2/3,-1/3)$ with only $u,d$  favors and $Q_e=Diag(2/3,-1/3,-1/3)$ with $u,d,s$  favors.
The corresponding vector and axial currents are given by: $j_V^\m=\jb Q_V\g^\m\j$, $j_A^\m=\jb  Q_A \g^\m\g_5\j$. In the linear response regime, the usual electric conductivity and the CESE conductivity  are defined by the generation of these currents upon an externally applied electric field via:
\begin{eqnarray}
\vec J_V&=&\s (e \vec E),\\
\vec J_A&=&\s_{\chi e} (e \vec E),
\end{eqnarray}
where $\vec E$ is the external electric field. These conductivities  can be computed by the Kubo's formulae:
\begin{eqnarray}
\s&=&\lim_{\o\ra 0}\lim_{\bk\ra0}\frac{i}{3\o}G_{Ve}^{R\, ii}(\o,\bk),\\
\s_{\chi e}&=&\lim_{\o\ra 0}\lim_{\bk\ra0}\frac{i}{3\o}G_{Ae}^{R\, ii}(\o,\bk),
\end{eqnarray}
where repeated $ii$ upper indices are summed over $1,2,3$. The $G_{\{V/A\} \,e}^{Rij}(x-y)=-i\h(x_0-y_0)\lan[J^i_{\{V/A\}}(x),J^j_e(y)]\ran$ are the retarded current-current correlation functions with $j_e^\m=e\jb Q_e\g^\m\j$ being the electric current, to be obtained via  analytic continuation from their counterparts in the
imaginary time formalism, $G^R(\o,\bk)=G(i\o_n\ra\o+i0^+)$. Diagrammatically the correlators
$G_{Ve}^{ij}$ and $G_{Ae}^{ij}$  are represented as in \fig{diag:kubo}, to be computed from the following expressions:
\begin{eqnarray}
G_{Ve}^{ij}(K)&=&-\sum_{a,b=1}^{N_c}\int_P\Tr_{\rm D,f}\ls Q_e\G^i_{ab}(P+K,P)S(P)\g^j \d_{ab}Q_V S(P+K)\rs,\\
G_{Ae}^{ij}(K)&=&-\sum_{a,b=1}^{N_c}\int_P\Tr_{\rm D,f}\ls Q_e\G^i_{ab}(P+K,P)S(P)\g^j\g_5 \d_{ab}Q_A S(P+K)\rs,
\end{eqnarray}
where the trace is over Dirac and flavor spaces, the integration $\int_P$ is given by $iT\sum_n\int d^3\bp/(2\p)^3$ with $p_0=i(2n+1)\p T$ the fermionc Matsubara frequency, and $S(P)$ is the quark propagator.

\begin{figure}[!htb]
\begin{center}
\includegraphics[width=13cm]{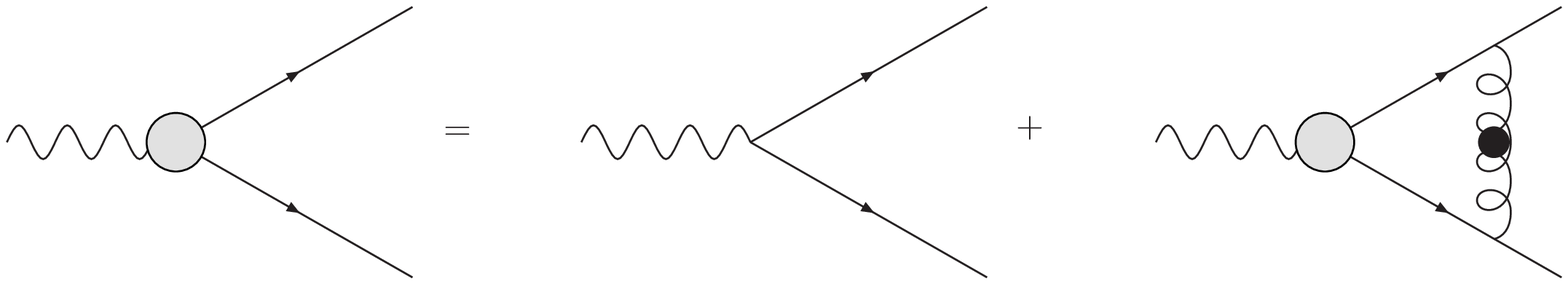}
\caption{Integral equation for the effective vertex $\G^\m$.} \label{diag:vertex}
\end{center}
\end{figure}

In the above formulation, what is nontrivial is the effective vertex $\Gamma$ for which a HTL resummation is need to cure the well-known pinch singularity (see e.g.~\cite{Jeon:1994if,Jeon:1995zm,MartinezResco:2000pz,ValleBasagoiti:2002ir,Aarts:2002tn}).  This is diagrammatically shown in \fig{diag:vertex}, and the vertex is to be solved from the following Bethe-Salpeter type equation:
\begin{eqnarray}
\G^\m_{ab}(P+K,P)&=&\g^\m\d_{ab}+g^2\sum_{c,d=1}^{N_c}\sum_{\a,\b=1}^{N_c^2-1}\int_Q\g^\r T^\a_{ac} S(P+K+Q)\non&&\times\G^\m_{cd}(P+K+Q,P+Q)S(P+Q)\g^\s T^\b_{db} \d^{\a\b}{^* D_{\r\s}}(Q).
\end{eqnarray}
In the above expression, $\int_Q=iT\sum_n\int d^3\bq/(2\p)^3$ with $q_0=i2n\p T$ being the bosonic frequency, $^*D_{\r\s}(Q)$  is the HTL propagator for gluons (with the left superscripted asterisk denoting a HTL resummed quantity). In the following we use a very similar procedure to that in~\cite{ValleBasagoiti:2002ir,Aarts:2002tn}. Clearly  $\G$ has a trivial color structure $\G^\m_{ab} = \G^\m \d_{ab}$, and we further define $\G^\m_{\pm}(\vec p) = \G^\m(p_0^s=\pm E_p+i0^+,\bp;p_0^s=\pm E_p-i0^+,\bp)$ with $E_p=p=|\vec p|$. Let us then introduce the contracted vertex functions (in the relevant kinematic limit):
\begin{eqnarray}
{\cal D}_+^s(\bp)=(p^i/2p^2)\bar{u}_s(\bp)\G^i_+(\bp)u_s(\bp)\, , \,\, {\cal D}_-^s(\bp)= (p^i/2p^2) \bar{v}_{-s}(-\bp)\G^i_-(\bp)v_{-s}(-\bp) \, .
\end{eqnarray}
where $s=\pm 1$ is the chirality index corresponding to right- and left-hand quarks.

\begin{figure}[!htb]
\begin{center}
\includegraphics[width=6cm]{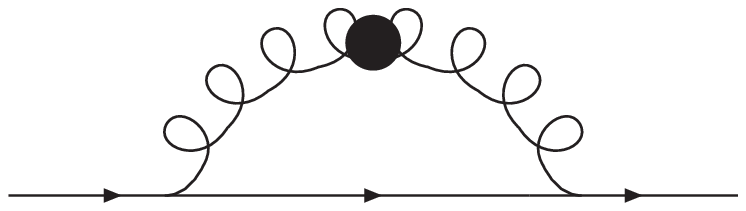}
\includegraphics[width=6cm]{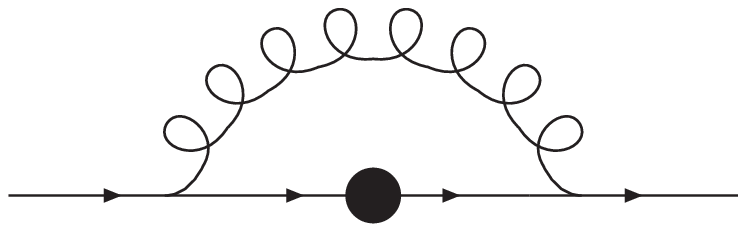}
\caption{One-loop self-energy  with one internal soft gluon (left) or one internal soft quark (right). } \label{diag:sf}
\end{center}
\end{figure}

With the above, we can re-express the Kubo's formulae as
\begin{eqnarray} \label{eq_conductivities_1a}
\s&=&- \frac{N_c}{3}\sum_s\Tr_{\rm f} Q_eQ_V\int\frac{d^3\bp}{(2\p)^3}\lc
n_F'(E_p-\m_s)\frac{{\cal D}_+^s(\bp)}{{\cal M}_+^s(\bp)}-n_F'(E_p+\m_s)\frac{{\cal D}_-^s(\bp)}{{\cal M}_-^s(\bp)}\rc,\\
\s_{\chi e}&=&- \frac{N_c}{3}\sum_ss\Tr_{\rm f} Q_eQ_A\int\frac{d^3\bp}{(2\p)^3}\lc
n_F'(E_p-\m_s)\frac{{\cal D}_+^s(\bp)}{{\cal M}_+^s(\bp)}-n_F'(E_p+\m_s)\frac{{\cal D}_-^s(\bp)}{{\cal M}_-^s(\bp)}\rc,
\label{eq_conductivities_1b}
\end{eqnarray}
where $n_F$ is the Fermi-Dirac distribution and $\mu_s\equiv s \mu_5$. We see the emergence of ${\cal M}_{\pm}^s$ which are the {\it thermal width} for quarks (``+'') and anti-quarks (``-'') with given chirality $s=\pm 1$. Such thermal width from the imaginary part of the (hard) quark/anti-quark ons-shell self-energy arises at one-loop level with one  internal soft gluon or one internal soft quark: see the diagrams in \fig{diag:sf}. They play the essential role of ``screening'' out the pinch singularity (as is evident in the ${\cal M} \to 0$ limit).    While in normal circumstance such thermal width would be identical for quarks/anti-quarks with any chirality, they become different in an environment with nonzero $\mu$ and $\mu_5$. The calculation of thermal width can be done in the standard HTL procedure (see e.g.~\cite{ValleBasagoiti:2002ir,Aarts:2002tn}), albeit generalized to include nontrivial impact from nonzero chemical potentials. For example, the Debye screening mass is now given by
\begin{eqnarray}
m_D^2&=&\frac{g^2}{3}\ls N_c T^2+\frac{1}{4}\Tr_{\rm f}\sum_{s=\pm}\lb T^2+\frac{3\m_s^2}{\p^2}\rb\rs.
\end{eqnarray}
With nonzero $\mu_5$ (that is, a P and CP odd environment), there is also a nontrivial, parity-odd contribution in the gluon self-energy $\Pi_A(P) \propto g^2 \mu_5 p (...)$. Such a contribution arises actually from VVA (two V for coupling to gluons and A for coupling to external $\mu_5$) triangle diagram (which is the origin of chiral anomaly).  As discussed in e.g.~\cite{Son:2012zy,Akamatsu:2013pjd}, this parity-odd term reproduces the chiral magnetic conductivity in proper kinematic limit. Such a term, however, can be neglected in the kinematic regime relevant to the present HTL computation of electric and CESE conductivities. The final results of ${\cal M}_{\pm}^s$ are similar with that in e.g.~\cite{Aarts:2002tn} but with some important modifications: There should be an additional Casimir factor  $C_F=(N_c^2-1)/(2N_c)$ accounting for the non-abelean nature, all the Fermi-Dirac distribution functions should now contain the chemical potential $\pm\m_s$, and finally, Eq.(3.22) of Ref.~\cite{Aarts:2002tn} is replaced by ${\cal M}_{\pm}^{s,sf,lo}=2\a_sC_F m_s^2\ln(1/g)p^{-1}[n_B(p)+n_F(\pm\m_s)]$ where the quark mass square
is given by
\begin{eqnarray}
m_s^2&=&\frac{g^2 C_F}{8}\lb T^2+\frac{\m_s^2}{\p^2}\rb.
\end{eqnarray}

At this point, we notice that the conductivities in Eqs.(\ref{eq_conductivities_1a},\ref{eq_conductivities_1b}) depend only upon the ratios ${\cal D}_{a}^s(\bp)/{\cal M}_{a}^s$ (with $a=\pm$ corresponding to quark/anti-quark and $s=\pm$ the chirality). Let us introduce the following dimensionless ratio functions:
\begin{eqnarray}
\f_a^s\lb x,\n_s\rb&=&\frac{a\, C_F}{4\pi}\,  g^4 \ln(1/g)  \, \frac{{\cal D}_{a}^s(\bp)}{{\cal M}_{a}^s/T}
\end{eqnarray}
with $x\equiv p/T$, $\n_s\equiv \m_s/T$. Starting from the effective vertex equation and following the steps as that in Refs.~\cite{ValleBasagoiti:2002ir,Aarts:2002tn} one can eventually reduce it to the following differential equation for the function $\f_a^s$:
\begin{eqnarray}
\label{diffeq}
1&=&\frac{2 \tilde{m}_s^2}{x}[n_B(x)+n_F(a\n_s)]\f_a^s(x)\non&&+
\frac{\tilde{m}_D^2}{x^2}\ls\f_a^s(x)-\lb1-\frac{x}{2}(1-2n_F[x-a\n_s])\rb x\frac{d\f_a^s(x)}{dx}-\frac{x^2}{2}\frac{d^2\f_a^s(x)}{dx^2}\rs,
\end{eqnarray}
with $\tilde{m}_D^2=m_D^2/(gT)^2, \tilde{m}_s^2=\left[(g^2C_F/8)(T^2+\mu_s^2/\pi^2)\right]/(gT)^2$. A most efficient way of solving this type of equation is a variational procedure, which has been previously employed in similar transport coefficient calculations~\cite{Arnold:2000dr,Arnold:2003zc,Dobado:2001jf,Aarts:2004sd,Chen:2006iga,Chen:2011km} and will also be used here to numerically solve $\f_a^s$. We address briefly this variational method in Appendix \ref{variation}. Once obtained, it can then lead straightforwardly to the final results for the conductivities:
\begin{eqnarray}
\s&=& [\Tr_{\rm f} Q_eQ_V]\, \sum_{s,a=\pm}  \s_{sa},\\
\s_{\chi e}&=&   [\Tr_{\rm f} Q_eQ_A]\,  \sum_{s,a=\pm}s \s_{sa},
\end{eqnarray}
where the species-specific contribution $\s_{sa}$ is computed from $\f_a^s$ through
\begin{eqnarray}
\s_{sa}&=&-a\frac{N_c}{3}\int\frac{d^3\bp}{(2\p)^3}n_F'(p-a\m_s)\c_a^s(p) \nonumber \\
&=& -\frac{2N_c}{3\p}\frac{T}{C_Fg^4\ln(1/g)}\int_0^\infty dxx^2n_F'(x-a\n_s)\f_a^s(x). \qquad
\end{eqnarray}

We have henceforth determined the CESE conductivity for high $T$ QGP within the HTL framework to the leading-log accuracy. To give a more concrete idea of actual numbers, let us consider two specific situations: \\
(1) With two light flavors $u,d$, $Q_e=Diag(2/3,-1/3)$, one obtains the following results
\begin{eqnarray}
\s^{2_f}&=&T\frac{\Tr_{\rm f}Q_eQ_V}{g^4\ln(1/g)}\lb 12.9989 + 6.50554 \frac{\m^2+\m_5^2}{T^2}\rb,\\
\s^{2_f}_{\chi e}&=&T\frac{\Tr_{\rm f}Q_eQ_A}{g^4\ln(1/g)}14.5163 \frac{\m\m_5}{T^2}.
\end{eqnarray}
\\
(2) With three light flavors $u,d,s$, $Q_e=Diag(2/3,-1/3,-1/3)$, one obtains the following results
\begin{eqnarray}
\s^{3_f}&=&T\frac{\Tr_{\rm f}Q_eQ_V}{g^4\ln(1/g)}\lb 11.8687  + 5.60215  \frac{\m^2+\m_5^2}{T^2}\rb,\\
\s^{3_f}_{\chi e}&=&T\frac{\Tr_{\rm f}Q_eQ_A}{g^4\ln(1/g)}13.0859  \frac{\m\m_5}{T^2}.
\end{eqnarray}
It shall be noted that the above concrete results in both cases confirm the predicted structures of these conductivities in Eqs.(\ref{eq_conductivities_kinetic1},\ref{eq_conductivities_kinetic2}) based on generic transport arguments. In a temperature regime of relevance to relativistic heavy ion collisions at RHIC and the LHC, one may expect strange quark mass to be non-negligible and the two flavor results may be more realistic.

\section{Summary}

In summary, we have determined a new transport coefficient for the quark-gluon plasma at very high temperature. This new coefficient $\sigma_{\chi e}$, the CESE (Chiral Electric Separation Effect) conductivity, quantifies the amount of axial current $\vec J_A$ that is generated in response to an externally applied electric field $e\vec E$: $\vec J_A = \s_{\chi e} (e\vec E)$. Starting with rather general argument in the kinetic theory framework, a characteristic structure $\s_{\chi e}\propto \mu \mu_5$ is identified, which is then confirmed by explicit results from the weak-coupling calculation. The perturbative result is first obtained here via Kubo's formulae within the HTL framework: $\s_{\chi e} = (\#) T\frac{\Tr_{\rm f}Q_eQ_A}{g^4\ln(1/g)}  \frac{\m\m_5}{T^2}$.  It is worth emphasizing again the CESE is an anomalous transport effect that occurs in a parity-odd environment with nonzero axial charge density (or equivalently axial chemical potential).

  It is also tempting to compare the perturbative results here with the strong coupling results obtained in certain holographic model~\cite{Pu:2014cwa} where it is found that $\s_{\chi e} \simeq T (\#) (8N_c\lambda/81)(\mu \mu_5/T^2)$ with the coefficient $\#\sim 0.025$. Comparing  numbers may not carry much meaning as the holographic model does not bear a direct correspondence to QGP. There are though a couple of remarkable features from this comparison: first, they share the same generic structural dependence $\s_{\chi e}\sim \mu \mu_5$; and second they have completely opposite dependence on coupling constant --- the HTL result gives $\s_{\chi e} \sim 1/[g^4\ln(1/g)]$  while the holographic result gives $\s_{\chi e} \sim \lambda \sim g^2$.

Let us end with discussions pertinent to phenomenology of high energy nuclear collisions. Firstly, it would be of great interest to extend such calculations to the temperature regime of few hundred $\rm MeV$, by e.g. employing certain nonperturbative effective models of QGP at those temperatures. Secondly, it would be very useful to develop realistic simulations to estimate the axial charge transport and possible azimuthal charge distribution as signal in e.g. Cu + Au collisions at RHIC, see the discussion in Ref.~\cite{Huang:2013iia}. The special distribution may be measured by using, e.g., the two particle correlation $\zeta_{q_1q_2}\equiv\lan\cos[2(\f_1+\f_2-2\j_{RP})]\ran$. Observation of a nonzero $\zeta_{++}+\zeta_{--}-2\zeta_{+-}$ may indicate the occurrence of CESE. Lastly, one may wonder about possible analogy of CESE in the case of a fluid field (rather than the electric field): a plausible guess would be $\vec J_A \propto   \mu\mu_5 \partial_t \vec v$ with $\vec v$ the flow velocity but more investigation is needed to tell whether this is the case. We leave these for future study to be reported elsewhere.

\vskip0.2cm

{\bf Acknowledgments---}
We thank D. Kharzeev, S. Lin, S. Pu, A. Tang, G. Wang, Q. Wang, D. Yang, H. Yee, and Y. Yin  for discussions and communications.  The research of YJ and JL is supported by the National Science Foundation (Grant No. PHY-1352368). The research of XGH is supported by Fudan University (Grant No. EZH1512519) and Shanghai Natural Science Foundation (Grant No. 14ZR1403000). JL is also grateful to the RIKEN BNL Research Center for partial support.

\appendix
\section{On the dissipative property of CESE}\label{diss}
In this Appendix we discuss the dissipative property of the CESE. To this end, we follow the discussions in Refs.~\cite{Huang:2009ue,Huang:2011dc} to examine the entropy production in the hydrodynamical framework. Let us write down the total vector and axial currents (for simplicity, we consider them both to be $U(1)$ currents) in the 4-vector form:
\begin{eqnarray}
{J^\mu_{V,tot}}&=&n u^\mu+J_V^\m,\non
{J^\mu_{A,tot}}&=&n_5 u^\mu+J_A^\m,\nonumber
\end{eqnarray}
where $n$ and $n_5$ are the vector and axial vector densities, $u^\m$ is the fluid velocity. The space-like vector  $J^\m_{V}$ and axial vector $J^\m_A$ contain higher-order terms in spatial derivative of velocity, external gauge potentials, etc; they are responsible to dissipative phenomena. Without external electromagnetic field (so that the $U(1)$ anomaly is turned off), both $J^\mu_{V,tot}$ and $J^\mu_{A,tot}$ are conserved, $\pt_\m J^\mu_{V/A,tot}=0$. This gives
\begin{eqnarray}
\pt_\m J^\m_V=-n\h-D n,\\
\pt_\m J^\m_A=-n_5\h-D n_5,
\end{eqnarray}
where $\h\equiv\pt_\m u^\m$ and $D\equiv u^\m \pt_\m$. In addition, the energy-momentum conservation, $\pt_\m T^{\m\n}=0$, gives
\begin{eqnarray}
D\ve+(\ve+P)\h=-J_V^\m (eE_\m),
\end{eqnarray}
where we neglect viscous corrections in $T^{\m\n}$ (which are not relevant to our discussions of currents). The $\ve$ and $P$ are the energy density and pressure, respectively. The entropy current can be written as
\begin{eqnarray}
s^\m=su^\m-\a J_V^\m-a_5 J_A^\m,
\end{eqnarray}
with $\a$ and $\a_5$ being two adjustable thermodynamical functions. A straightforward calculation of the divergence of $s^\m$ gives the entropy production
\begin{eqnarray}
T\pt_\m s^\m=-(T\a-\m)\pt_\m J_V^\m-(T\a_5-\m_5)\pt_\m J_A^\m-J_V^\m(eE_\m+T\nabla_\m\a)-J_A^\m T\nabla_\m\a_5,
\end{eqnarray}
where $\nabla_\m\equiv (g_{\m\n}-u_\m u_\n)\pt^\m$ and we have used the thermodynamical relations
\begin{eqnarray}
\ve+P&=&Ts+\m n+\m_5 n_5,\non
D\ve&=&TD s+\m D n+\m_5 Dn_5.\nonumber
\end{eqnarray}
The second law of thermodynamics requires that $T\pt_\m s^\m\geq 0$, which enforces the conditions $\a=\m/T$ and $\a_5=\m_5/T$. From these conditions, one can then obtain
\begin{eqnarray}
T\pt_\m s^\m=-J_V^\m(eE_\m+T\nabla_\m\a)-J_A^\m T\nabla_\m\a_5.
\end{eqnarray}
Next we substitute into the above the following relations which hold to the first order in the derivative expansion:
\begin{eqnarray}
J_V^\m=\s (eE^\m+T\nabla^\m\a)+T\l\nabla^\m\a_5,\\
J_A^\m=\s_{\c e} (eE^\m+T\nabla^\m\a)+T\l_{\c5}\nabla^\m\a_5,
\end{eqnarray}
where $\s$ is the usual electric conductivity, $\s_{\c e}$ is the CESE conductivity, $\l$ and $\l_{\c 5}$ are two new transport coefficients that characterize how the system responds to spatial gradient of $\a_5$ (or equivalently, to an applied axial electric field $e\vec E_5$). The result is
\begin{eqnarray}
T\pt_\m s^\m=-\s(eE_\m+T\nabla_\m\a)(eE^\m+T\nabla^\m\a)-\l_{\c 5}T^2\nabla^\m\a_5\nabla_\m\a_5-T(\s_{\c e}+\l)\nabla_\m\a_5(eE^\m+T\nabla^\m\a).\nonumber
\end{eqnarray}
The non-negativeness of $T\pt_\m s^\m$ requires that $\s\geq0$ and $\l_{\c 5}\geq0$ and $\s\l_{\c 5}\geq (\l+\s_{\c e})^2/4$. Furthermore, we have $\s_{\c e}=\l$ owing to the Onsager reciprocal principle, and thus the last inequality can be written as
\begin{eqnarray}
\s\l_{\c 5}\geq \s_{\c e}^2,
\end{eqnarray}
which means the CESE disappears when either the usual conduction $\s$ or $\l_{\c 5}$ vanishes. It is also evident that the above constraint from entropy production can not uniquely fix the sign of $\s_{\c e}$.

Thus we conclude that the CESE does affect the entropy production and in this sense it is dissipative in nature. However, the CESE appears always in accompany with the usual conduction $\s$ and $\l_{\c 5}$, It does not generate entropy by itself, under either an applied electric field $e\vec E$ or an axial electric field $e\vec E_5$: This is why the CESE conductivity can be either positive or negative.

\section{Variational procedure for solving Eq.~(\ref{diffeq})}\label{variation}
An elegant and very efficient way to solve \eq{diffeq} is the variational method. The starting point is the following functional of $\f$
\begin{eqnarray}
Q_{a}^s[\f]
&=&\int_0^\infty dx \big\{x^2n_F(x-a\n_s)[1-n_F(x-a\n_s)]\f(x)\non&&-\tilde{m}_s^2 xn_F(x-a\n_s)[1-n_F(x-a\n_s)][n_F(a\n_s)+n_B(x)]\f^2(x)
\non&&-\frac{\tilde{m}^2_D}{4}n_F(x-a\n_s)[1-n_F(x-a\n_s)][x^2(\f'(x))^2+2\f^2(x)]\big\}\non
&\equiv&\lan\f|S\ran-\frac{1}{2}\lan\f|C|\f\ran.
\end{eqnarray}
It can be shown by direct calculation that $\d Q_a^s/\d\f(x)$ leads to \eq{diffeq} under the condition $\lim_{x\ra0}\f(x)<\infty$, and
\begin{eqnarray}
\s_{sa}&=&
\frac{4N_c}{3\p}\frac{T}{C_Fg^4\ln(1/g)}Q_a^s\big|_{\rm max},
\end{eqnarray}
where $Q_a^s\big|_{\rm max}$ is $Q_a^s[\f]$ with $\f$ satisfying \eq{diffeq}.
Suppose $\{B_i|i=1,2,\cdots,n\}$ is a set of basis with elements satisfying
\begin{eqnarray}
\label{orthognal}
\lan B_i|C|B_j\ran&\equiv&\int_0^\infty dx \big\{2\tilde{m}_s^2 xn_F(x-a\n_s)[1-n_F(x-a\n_s)][n_F(a\n_s)+n_B(x)]B_i(x)B_j(x)
\non&&+\frac{\tilde{m}^2_D}{2}n_F(x-a\n_s)[1-n_F(x-a\n_s)][x^2B_i'(x)B_j'(x)+2B_i(x)B_j(x)]\big\}\non
&=&\d_{ij}.
\end{eqnarray}
Write $\f(x)=\sum_{i=1}^n a_iB_i(x)$. Thus,
\begin{eqnarray}
Q_{a}^s[\f]
&=&\sum_{i=1}^n a_i\lan B_i|S\ran-\frac{1}{2}\sum_{i,j=1}^n a_ia_j\lan B_i|C|B_j\ran\non
&=&\sum_{i=1}^n \lb a_i\lan B_i|S\ran-\frac{1}{2}a_i^2\rb.
\end{eqnarray}
The stationary condition of $Q_a^s[\f]$ is now
\begin{eqnarray}
a_i=\lan B_i|S\ran,
\end{eqnarray}
and
\begin{eqnarray}
Q_{a}^s\big|_{\rm max}=
\frac{1}{2}\sum_{i=1}^na_i^2.
\end{eqnarray}
In practice, we choose $B_i(x)$ as a polynomial up to powers of $x^i$ (particularly, $B_0$ is chosen to be independent of $x$ and its value is given by the normalization (\ref{orthognal})) and once $B_i(x)$ ($i=0,1,\cdots, n$) is known we determine $B_{n+1}$ by imposing \eq{orthognal}.
This variational method converges very fast, only first a few $B_i$'s can lead to very accurate results. We checked that
using first two and using first six polynomials leads to relative error within $10^{-4}$. The results presented in the main text are obtained by using the first six polynomials.

 \vfil

\end{document}